\documentclass[showpacs,prl,twocolumn]{revtex4}

\usepackage{graphicx,graphics}
\usepackage{supertabular}%
\usepackage{bm}%
\usepackage{amsmath}%

\begin{document}

\title{Femtosecond superradiance in multiple-section
InGaN/GaN quantum well laser structures}

\author{D. L. Boiko}
\email{dmitri.boiko@csem.ch} \affiliation{CSEM Centre Suisse d'Electronique et de Microtechnique, 2001 Neuchatel, Switzerland}
\author{P P Vasil'ev}
\affiliation{P N. Lebedev Physical Institute, 53 Leninsky prospect, Moscow 119991, Russia}
\date{ \today }

\begin{abstract}
We analyze theoretically a possibility of the superradiant emission (SR) in GaN-based quantum well (QW) laser heterostructures. Our model is based on  travelling wave Maxwell-Bloch equations predicts building up of macroscopic coherences in the ensemble of carriers in the QWs. We show that SR is covered by the Ginzburg-Landau equation for phase transition to coherent matter state. The generation of superradiant pulses as short as 500 fs at peak powers of over 200 W has been predicted for InGaN/GaN heterostructure QWs with the peak emission in the blue/violet wavelength range.
\end{abstract}


\maketitle








A spontaneous build-up of macroscopic coherences in solids, or in other words, a phase transition to a single macroscopic quantum state of matter, has always been a fascinating subject for scientific and engineering communities. Recent examples of quantum coherent phenomena include the Bose-Einstein condensation (BEC) of atoms, which stipulated works on laser-cooled atomic clocks with unprecedented fractional frequency stability, and wide spreading use of superconducting magnets in which a coil undergoes the BCS (Bardeen, Cooper, and Schrieffer) transition.
The search for spontaneous build-up of macroscopic coherences in semiconductor microcavities has resulted in intensive research towards a polariton laser. Interestingly, group III - nitride semiconductors
are capable of reaching macroscopic coherences at room temperature conditions. Thus due to large reduced exciton mass, the InGaN/GaN quantum well (QW) exciton binding energy is higher as compared to GaAs/AlGaAs QWs or other conventional III-V counterparts. Furthermore, due to a stronger exciton-photon Rabi coupling, InGaN/GaN microcavities offer possibility of polariton lasing at room temperature conditions \cite{Christopoulos07}. High critical temperature of BEC transition is conditioned by 
low in-plane effective mass of 
polaritons, so as the critical density is reached before destruction of excitons. Such 2D BEC is a transient dynamic state, which can nevertheless be analyzed using a stationary Schr\"{o}dinger equation \cite{Boiko08}.

The superradiance (SR) in semiconductor edge emitters can be considered as 
another example of spontaneous macroscopic coherences in solids \cite{Vasil'ev09}.
The cooperative radiative recombination in an ensemble of quantum oscillators (e.g. atoms or molecules) 
has been predicted before the invention of lasers \cite{Dicke54}. Since that, it has been extensively studied both theoretically and experimentally \cite{Skribanowitz73,Schuurmans81}. The characteristic features of the SR emission are the temporal and spatial coherence, highly anisotropic emission pattern, quadratic dependence of pulse intensity on the number of excited atoms $I \propto n^2 $, 
afterpulse ringing attributed to Rabi-type oscillations. The SR pulse duration decreases with the number of emitters $\tau_{sr} \propto 1/ n$, while the pulse energy is proportional to the ensemble population ($I\tau_{sr}\propto n $), in agreement with the energy  conservation considerations. On the other side, the spontaneous nature of the transition to the transient macroscopically coherent state is responsible for large fluctuations in the shape, duration and amplitude of SR pulses due to quantum-mechanical uncertainties 
\cite{Gross82,Men'shikov99}.

In semiconductors, a hypothesis has been drawn that the SR is assisted by formation of a transient coupled electron-hole (e-h) pair state mediated by photons, much like as phonon mediation in a 
superconducting BCS state \cite{Vasil'ev01,Vasil'ev04}. 
According to that  the e-h system undergoes the second order non-equilibrium phase transition when the coherent e-h BSC-like state is building up during SR pulse emission. From practical point of view, SR has been considered as one of the promising approaches to generate high-power femtosecond optical pulses exceeding by a few orders of magnitude the power in lasing or amplified spontaneous emission regimes.

\begin {figure}[tbp]
\includegraphics {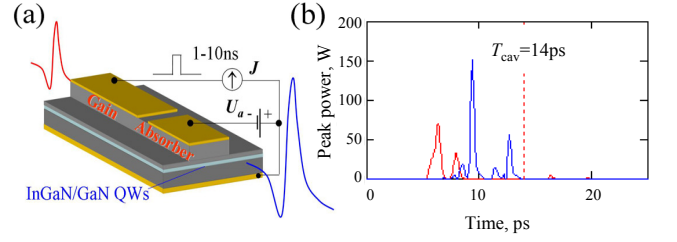} 
\caption {(a) Schematic of two-section laser structure for SR emission generation and (b) output SR pulses at the gain section facet (red curve) and absorber section facet (blue curve).}\label {fig2_Cavity}
\end {figure}

In this letter, inspired by remarkable features of the group III - nitride microcavities in reaching room temperature BEC macroscopic coherences, we studied numerically the Dicke superradiance in an edge-emitting ridge-waveguide multiple-section cavity with InGaN/GaN QWs (see Fig.\ref{fig2_Cavity}). We developed a numerical 
model based on a semi-classical 
description of macroscopic coherences (polarization) 
in the ensemble of electrons and holes
in the QWs. In distinguishing from previous theoretical treatment developed for the bulk and QW GaAs-based  materials \cite{Vasil'ev99}, 
our new numerical model takes correctly the carrier populations and coherences into account in function of the pump current and absorber bias. To support the hypothesis of BCS-like condensation, we show that the travelling wave Maxwell-Bloch equations governing the SR emission can be converted
to the Ginzburg-Landau equation for the second-order BCS phase transition. This  enables us to define the critical density in the system, the order parameter and the coherence time (and length).

As a model system we utilize a tandem cavity configuration consisting of two separately contacted sections (Fig. \ref{fig2_Cavity}). The shorter negatively biased section is used as a saturable absorber. The longer section is pumped with short current pulses and provides the optical gain. 

The evolution of the amplitudes of the forward ($A_+$) and backward ($A_-$) travelling waves 
is defined by the cavity loss and macroscopic medium polarization
\vspace{-0.05in}\begin{equation}
\frac{\partial A_\pm}{\partial t } \pm v_g\frac{\partial A_\pm}{\partial z }= \frac 1 2 \Gamma \sqrt{\frac {g_0}{T_2} }P_\pm-\frac 1 2 v_g \alpha_i A_\pm,
%
%
\label{TravAmpl}
\vspace{-0.05in}\end{equation}
where the field amplitudes are normalized in the secondary quantization convention $\hat{a}_\pm\left|N_\pm\right\rangle{=}A_\pm\left|N_\pm{-}1\right\rangle$:
\vspace{-0.1in}\begin{equation}
\textbf{E}_\pm{=}\sqrt{\frac{4 \pi \hbar \omega v_g^2}{c^2}} \textbf{e}_\pm A_\pm \sin(\omega t \mp kz),\quad A_\pm{=}\sqrt{N_\pm}
\label{NormA}
\vspace{-0.05in}\end{equation}
The 
normalization used for coherences (the ensemble average for the off-diagonal density matrix elements) induced between the electrons and holes and associated with the forward and backward waves reads:
\vspace{-0.05in}\begin{equation}
\textbf{P}_\pm{=}\sqrt{\frac{c^2 g_0 }{4 \pi \hbar \omega v_g^2 T_2}} \textbf{e}_\pm P_\pm \cos(\omega t \mp kz)
\label{NormP}
\vspace{-0.05in}\end{equation}
In convention (\ref{NormP}), the macroscopic variables $P_\pm$ measure the order parameter of the system. The usual boundary conditions are applied at left ($z{ =}0$) and right ($z {=} L$) cavity mirrors
\vspace{-0.05in}\begin{equation}
A_{+}(0,t){=}\sqrt R A_{-}(0,t),\quad A_{-}(L,t){=}\sqrt R A_{+}(L,t).
\label{BC}
\vspace{-0.05in}\end{equation}
Separate equations for the gain and absorber sections take into account that (i) the carrier lifetime in absorber is much shorter as compared to the gain; (ii) pumping by carrier injection is applied to the gain section only, (iii) Quantum Confined Stark Effect (QCSE) absorption tuning occurs in the absorber section.
The dynamics of carrier populations and coherences is described as follows:
\vspace{-0.05in}\begin{equation}
\hspace{-0.1in}\begin{split}
&\frac{\partial P_\pm}{\partial t }{=}{-}\frac{P_\pm}{T_2}{+}D\frac{\partial^2 P_\pm}{\partial z^2}{+}\sqrt{\frac{g_0}{T_2}}(n{-}n_t) A_\pm {+} \Lambda_\pm \\
&\frac{\partial n}{\partial t }{=}D\frac{\partial^2 n_\pm}{\partial z^2}{-}\frac{n}{\tau_n}{-}\sqrt{\frac{g_0}{T_2}}(A_{+}P_{+}{+}A_{-}P_{-}){+}\frac{J(z{,}t)}{e_q d}
\\
&\frac{\partial n_a}{\partial t }{=}D\frac{\partial^2 n_a}{\partial z^2}{-}\frac{n{-}n_V}{\tau_a}{-}\sigma\sqrt{\frac{g_0}{T_2}}(A_{+}P_{+}{+}A_{-}P_{-})
\label{P&n}
\end{split}
\vspace{-0.05in}\end{equation}
where the carrier diffusion D is taken into account; $\sigma$  is the differential absorption to gain ratio. The Langevin force term  due to polarization (thermal) noise $\Lambda_{\pm}$ triggers the spontaneous build-up of superradiance pulse.

\begin {figure}[tbp]
\includegraphics[width=7.2cm]{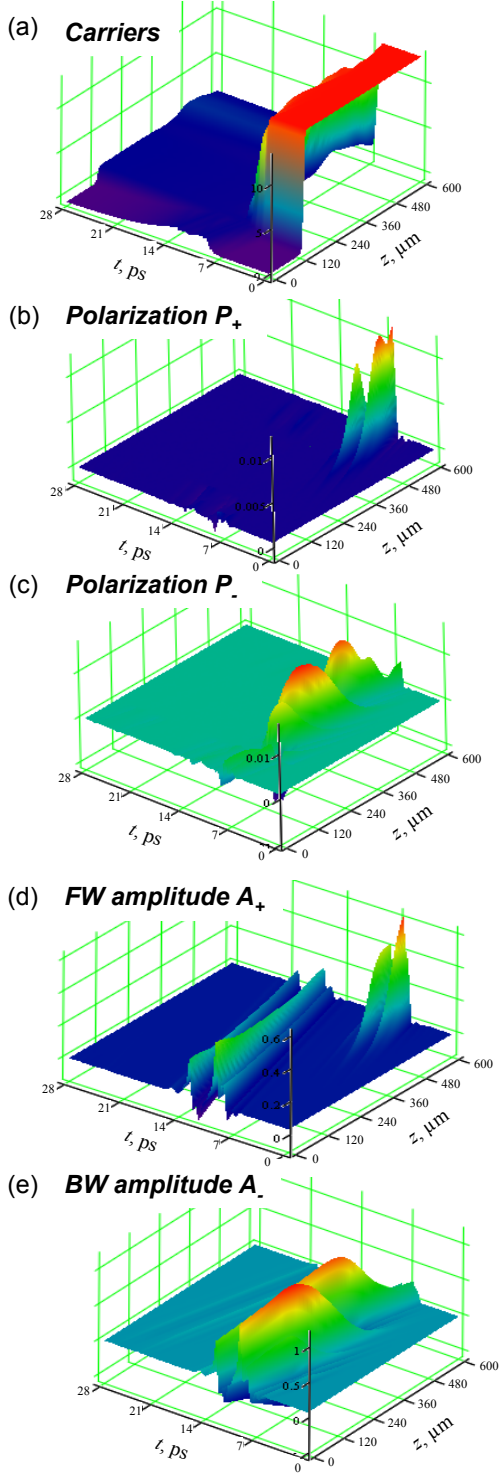}
\caption {Spatiotemporal dynamics of carrier populations (a), macroscopic polarization (carrier coherences) in InGaN/GaN QWs associated with the forward (b) and backward (c) waves, field amplitudes of the forward (d) and backward (e) waves. The axes $T$ and $Z$ are normalized on cavity roundtrip time and cavity length, respectively. The polarization and the field amplitudes are  normalized on saturation parameter.
The current amplitude I = 211 mA. 
}
\label {fig345_TZDyn}
\end {figure}

Figure \ref{fig345_TZDyn} displays the numerical solution of the above equations for the SR emission in the cavity of the overall length $L$=600$\mu$m and relative absorber length 20\%. The absorber section is located at the left facet of the structure (at $z{=}0$) and the gain section is situated at the right facet. 

Initially, a high reversed bias is applied to the absorber leading to depletion of carriers in the QWs  (blue color in the scale of Fig.\ref{fig345_TZDyn}(a)
). It prevents the structure from lasing and enables accumulation of carriers in the gain section. The gain section QWs are pumped to a high level, so as in this particular case, the carrier population is about 15 times above the transparency $n_t$ (red color in Fig.\ref{fig345_TZDyn}(a). The entire process starts at $t$=0 from the spontaneous polarization noise. 
The rate of spontaneous polarization noise into the cavity mode ${\sim} \frac 1 2 \Gamma \sqrt{g_0 T_2}  \Lambda_{\pm}$  
corresponds to 
27 photons per cavity round trip.
This cannot be seen in the scale of Fig.\ref{fig345_TZDyn}. 
The noise sources are uniformly distributed along the cavity 
but the macroscopic coherences  are building up 
at the 
edges of the gain section, 
after about half of the cavity roundtrip time ($T_\text{cav}$=14ps). Because of the asymmetric cavity,
the shape and emission time of the output SR pulses from the left and right cavity facets are different (Fig.\ref{fig2_Cavity}(b)) and the emitted SR pulses 
are not related by reflections at the cavity facets.

The build-up of the macroscopic polarization is followed by Rabi oscillations at a frequency dependent on the pumping rate . They are clearly seen in the
wave amplitudes (Figs.\ref{fig345_TZDyn}(d) and (e)). The carrier density drops abruptly at the same time when the polarization rises indicating that almost all carriers contribute to the field (Figs.\ref{fig345_TZDyn}(a) and (c)).  When the travelling backward pulse hits the absorber section at $z{=}0.2L$, the pulse intensity is sufficiently high to saturate  absorber 
(the blue step in the region $0{<}z{<}120\mu$m at $t{=}7ps$ in Fig.\ref{fig345_TZDyn}(a)). From this time, there is no absorption and gain in the cavity. The backward travelling SR pulse reflected of the left cavity facet can now freely travel through the structure yielding emission of the secondary pulse at the right cavity facet (Fig.\ref{fig2_Cavity}(b), red curve).  In this particular realization of SR emission, the pulse emitted form the absorber section side is shorter due to sharpening of its rising front in saturating absorber. 

\begin {figure}[tbp]
\includegraphics {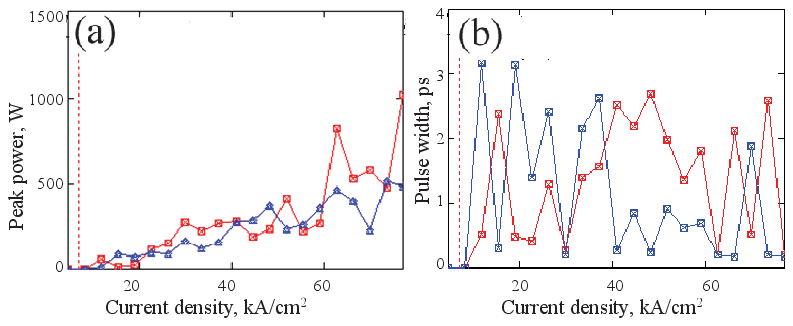}
\caption {SR  pulse peak power (a), pulse width (b).
}\label {fig67_LI}
\end {figure}

The build-up of macroscopic polarization has a stochastic nature
, which results in differences between individual realizations of  SR pulses. The peak power and pulsewidth of SR pulses vary within the range of 50-200 W  and 0.5-5 ps, respectively, depending on the driving conditions
(Fig.\ref{fig67_LI}).
Varying the overall cavity length and the relative length of absorber, we find that the intensity ratio and the width 
of the forward and backward SR pulses can be effectively altered.
The common feature of all realizations 
is that the polarization and carrier excess above the transparency 
vanish 
after a few oscillations so as the 
the optical pulses 
travel freely 
without gain or absorption.

In Fig.\ref{fig345_TZDyn} (e),
the peak intensity of the pulse 
corresponds to photon density in the cavity of 9$\cdot$10$^{18}$ cm$^{-3}$. The density of the associated coherence excitations $P_{\pm}$ at the polarization peak maxima is of 2.8$\cdot$10$^{21}$ cm$^{-3}$, which is much higher than the initial carrier density $n_0$=2.4$\cdot$10$^{20}$  cm$^{-3}$ .
A hypothesis has been previously stated \cite{Vasil'ev01,Vasil'ev04} 
that the SR in a semiconductor is mediated by a formation of the transient
BCS-like state of e-h pairs.

In order to interpret the SR emission in terms of BCS-like transition one should transform Eqs.(\ref{TravAmpl}) and (\ref{P&n}) to the form of the Ginzburg-Landau equation (GLE) for the second-order BCS phase transition. The difficulty is caused by the fact the GLE is essentially a steady-state equation for an isolated system, while the SR is a transient process. To overcome this difficulty we change variables to the internal coordinates of the 
pulses. For simplicity we consider only the forward traveling pulse, introducing  new coordinate system  $\zeta{=}t-z/v_g$ and $z$. The phase transition under the question is related to the change of the effective phase relaxation time $T_2^{\text{eff}}$, which is comparable with the lifetime of photons ($1/v_g \alpha_i$=3ps). At the same time, the carrier relaxation can be neglected
at the time scale of the pulse width:
\vspace{-0.05in}\begin{equation}
\begin{split}
\frac{\partial n}{\partial \zeta}&{=}{-}\sqrt{\frac{g_0}{T_2}}A_{+} P_{+}, \, \frac{\partial A_+}{\partial z}{=} \frac{\Gamma}{2 v_g} \sqrt{\frac{g_0}{T_2}}
P_{+} {-}\frac {\alpha_i}2  A_+, \\
\frac{\partial P_{+}}{\partial \zeta}&{=}{-}\frac{P_+}{T_2}{+}\sqrt{\frac{g_0}{T_2}}(n{-}n_t)A_+.
\label{Space1}
\vspace{-0.5in}\end{split}
\end{equation}
The analytical solution is 
obtained for a model with localized parameters.
The substitution $1/L^* \rightarrow \partial/ \partial z$ yields $A_+{=}P_+ \Gamma \sqrt{g_0/T_2}/ v_g(\alpha_i{+}2/L^*) $ with $L^*$ being some effective length (see the discussion below), which allows us to exclude  the wave amplitude:
\vspace{-0.05in}\begin{equation}
\begin{split}
\frac{\partial P_+}{\partial \zeta}&{=}\frac{\Gamma g_0 (n{-}n_{cr})}{ T_2 v_g (\alpha_i {+} 2{/}L^*) }P_+, \; n_{cr}{=}n_t{+}\frac{v_g(\alpha_i{+}2/L^*)}{\Gamma g_0},\\
\frac{\partial n}{\partial \zeta}&{=}{-}\frac{\Gamma g_0 P_+^2}{v_g T_2 (\alpha_i{+}2/L^*)}{=}{-}\frac{P_+}{(n{-}n_{cr})}\frac{\partial P_+}{\partial \zeta},
\label{Space2}
\end{split}
\vspace{-0.1in}\end{equation}
The last equation recovers the Bloch vector conservation $(n(\zeta){-}n_{cr})^2{+}P_{+}(\zeta)^2{=}(n_0{-}n_{cr})^2 $ with $ n_0=n(- \infty) $ being the initial carrier density in the system. Finally, taking the second derivative in the first equation (\ref{Space2}) one obtains the master equation:
\vspace{-0.05in}\begin{equation}
{-}\frac{ \partial^2 P_+}{\partial \zeta^2 }{+}
\frac{\Gamma^2 g_0^2 {L^*}^2 }{T_2^2 v_g^2(2{+}\alpha_i L^*)^2}\Bigl[(n_0{-}n_{cr})^2
{-} 2 P_+^2 \Bigr]P_+{=}0.
\label{GLE}
\vspace{-0.05in}\end{equation}
This equation has a canonical form of the Ginzburg-Landau equation (GLE), in which     $P_+$ is the order parameter of the system and the internal coordinate $\zeta{=}t{-}z/v_g$ plays the role of spatial variable. For the carrier densities above the critical $n_{cr}$ (see Eq.(\ref{Space2})), its steady state solution defines the condensate fraction (BCS-like). The analytical similarities between Eq.(\ref{GLE}) and GLE, allows one to assume that for $n{>}n_{cr}$, the evolution of the the order parameter is defined by
hyperbolic secant function. Substituting  
$P_+{=}P_0\text{sech}(\zeta/\tau_p)$, we obtain the coherence time $\tau_c{=}2\ln(1{+}\sqrt{2})\tau_p$ (the FWHM of the SR pulse), and the peak photon number in the cavity ($N_{+}{=}A_{+}^2(0)$)
\vspace{-0.1in}\begin{equation}
\tau_p{=}\frac{ T_2 v_g {(}\alpha_i {+} 2/L^*{)}}{\Gamma g_0 (n_0{-}n_{cr})}, \quad
N_{+}{=}
\frac{T_2}{g_0\tau_p^2}\propto(n_0{-}n_{cr})^2
\label{Tcoh}
\vspace{-0.1in}\end{equation}
The obtained pulse width $2\ln(1{+}\sqrt{2})\tau_p$ and peak power  $ \propto \hbar \omega v_g N_{+}$ exhibit the initially expected behavior for the SR pulse 
known from the studies of SR in a gas medium.
We thus shown that (i) the SR in semiconductor is the BCS-like phase transition and follows the GLE (\ref{GLE}) and (ii) the order parameter of the system showing the abrupt growth is the medium polarization $P_{\pm}$, which is defined by coherences (off-diagonal terms) in the density matrix. This indicates that the wavefucntion of the condensate quasi particles is the composed of products of the e and h wavefucntions.
The proposed theory of BCS-like condensation during the SR emission is different from all previous treatments. Although it does not involve the system temperature explicitly, the thermodynamic temperature can be introduce to characterize the the degree of population inversion, that is the initial quasi equilibrium state of carriers of density $n_0$.

The SR pulse features (\ref{Tcoh}) are in perfect agrement with the predictions of the Dicke theory if the coherence length $L_c{=}\tau_c v_g$ is longer than the cavity length $L$. In that case, the characteristic length of the condensate fraction $L^*$ is defined by the sample size  $L^*{=}L{<}L_c$.
However with increasing carrier density $n_0$, the coherence length $L_c$ becomes shorter than the cavity length $L$ and the size of the condensate fraction is set by the coherence length $L^*{=}L_c{<}L$. As can be seen from Fig. \ref{fig345_TZDyn} (b) and (c) showing the order parameter $P_{\pm}$, this is the typical situation in the semiconductor laser cavity. Thus for $n>n_{cr2}$ \cite{BoikoPrep}, Eq.(\ref{Tcoh}) leads to a quadratic equation with respect to $\tau_p$. Its  asymptotic behavior for $n\gg n_{cr2}$ reads
\vspace{-0.1in}\begin{equation}
\tau_p \propto \sqrt{\frac{ T_2}{\Gamma g_0 (n-n_t)}}, \quad
N_+ 
\propto(n_0{-}n_{t})^{3/2}
\label{Tcoh2}
\vspace{-0.1in}\end{equation}
It is interesting to compare the predictions of our simplified analytical model with the results of our numerical experiment in Fig.\ref{fig345_TZDyn}. The critical carrier density of $n_{cr}{=}2.52n_t$ is achieved at 4.2 kA/cm$^2$  pump current density ($n{=}\tau_n J/e_q d$) so as the condensation condition $n{>} n_{cr}$  is fulfilled. However at $n_{cr2}=2.57n_t$  (at $J$=4.3kA/cm$^2$) the coherence length reduces down to the size of the cavity. Thus only in the narrow range $n_{cr}<n<n_{cr2}$, the SR pulse peak power exhibits Dicke's quadratic growth (\ref{Tcoh}). At higher injection levels, it is switching to the dependence with the asymptotic (\ref{Tcoh2}). This feature is perfectly reproduced by the numerical model in Fig. \ref{fig67_LI}. In conditions of Fig.\ref{fig345_TZDyn}, the estimated pulse width  $1.76\tau_p$ is 600fs.

The L-I and pulse width curves in Fig.\ref{fig67_LI} exhibit large amplitude and timing instabilities of the output SR pulses. The SR is particularly interesting quantum optics phenomenon because quantum fluctuations manifest themselves macroscopically in the time and energy domains. The predicted macroscopically large fluctuations are one of the characteristic features of a condensed state \cite{Butov94}. 

In summary, we presented the  numerical model of Dicke SR in semiconductor laser cavities and shown that it corresponds to the analytical Ginzburg-Landau equation for BCS phase transition. We have demonstrated that the theoretically predicted performance of SR pulses in InGaN/GaN QW laser heterostructures 
exceeds substantially that of the mode-locked or Q-switched GaN-based lasers \cite{Kuramoto10}.

\begin{acknowledgments}
This research is supported by the EC Seventh Framework Programme FP7/2007-2013 under the Grant Agreement n° 238556 (FEMTOBLUE)
\end{acknowledgments}

\end{document}